\documentclass[journal]{IEEEtran}
\usepackage{graphicx}
\usepackage{amsmath,amssymb,amsfonts}
\usepackage{algorithmic}
\usepackage{array}
\usepackage{textcomp}
\usepackage{subfig}
\usepackage{cite}
\usepackage{url}

\begin{document}

\title{A Single-Point Measurement Framework for Robust Cyber-Attack Diagnosis in Smart Microgrids Using Dual Fractional-Order Feature Analysis}

\author{\IEEEauthorblockN{Author Name}
\IEEEauthorblockA{Department\\
Institution\\
Email: author@institution.edu}}

\maketitle

\begin{abstract}
Cyber-attacks jeopardize the safe operation of smart microgrids. At the same time, existing diagnostic methods either depend on expensive multi-point instrumentation or stringent modelling assumptions that are untenable under single-sensor constraints. This paper proposes a Fractional-Order Memory-Enhanced Attack-Diagnosis Scheme (FO-MADS) that achieves low-latency fault localisation and cyber-attack detection using only one VPQ (Voltage-Power-Reactive-power) sensor. FO-MADS first constructs a dual fractional-order feature library by jointly applying Caputo and Grünwald-Letnikov derivatives, thereby amplifying micro-perturbations and slow drifts in the VPQ signal. A two-stage hierarchical classifier then pinpoints the affected inverter and isolates the faulty IGBT switch, effectively alleviating class imbalance. Robustness is further strengthened through Progressive Memory-Replay Adversarial Training (PMR-AT), whose attack-aware loss is dynamically re-weighted via Online Hard Example Mining (OHEM) to prioritise the most challenging samples.

Experiments on a four-inverter microgrid testbed comprising 1 normal and 24 fault classes under four attack scenarios demonstrate diagnostic accuracies of 96.6\% (bias), 94.0\% (noise), 92.8\% (data replacement), and 95.7\% (replay), while sustaining 96.7\% under attack-free conditions. These results establish FO-MADS as a cost-effective and readily deployable solution that markedly enhances the cyber-physical resilience of smart microgrids.
\end{abstract}


\section{Introduction}
Power electronic converters in microgrids must be continuously monitored to guarantee safe operation. Recent studies have proposed noise-resilient residual networks [1], chaotic-DC arc-grounding detection [2], capsule-based compound-fault decoupling [3], and lightweight online classifiers for multiphase drives [4], illustrating the demand for robust data-driven monitoring. Open-circuit IGBT failures can bypass protection circuits and create persistent current imbalance and torque ripple [5]. Traditional model-based diagnosis depends on accurate parameters and uniform hardware, whereas the growing integration of information and communication technologies (ICT) for coordinating distributed energy resources (DERs) exposes microgrids (MGs) to new cyber-threats: cross-layer attacks degrade performance [6], and heavy network traffic increases vulnerability [7]. Limited situational awareness leaves MGs susceptible to false-data injection (FDI), denial-of-service (DoS), and other deception attacks.

Attack-resilient control strategies have therefore gained traction. Zhou \textit{et al.} [6] introduced a cross-layer distributed controller against FDI and DoS, while Yao \textit{et al.} [7] designed latency-tolerant adaptive control for islanded MGs. Shi \textit{et al.} [8] addressed deception in DC MGs, and Shen \textit{et al.} [9] employed event-triggered secondary control to curb communication burden under attack conditions. Complementing these defences, Xia \textit{et al.} [10] demonstrated data-driven online gain scheduling for time-delayed MGs, underscoring the promise of intelligent methods.

Parallel efforts focus on data-driven converter fault diagnosis. Machine-learning models map electrical waveforms to fault labels without explicit physical models; representative techniques include [1]--[4]. Their accuracy, however, deteriorates with low-quality or out-of-domain data. Robust methods such as the low-quality-tolerant approach of Xia and Xu [11] and the transferable models in [12], [13] mitigate these issues, while safe deep reinforcement learning has been explored for hierarchical MG control [14].

In practice, multiple inverters operate in parallel, requiring localisation of the faulty inverter before identifying the damaged switch, all under possible cyber-attacks. Zhou \textit{et al.} [15] proposed cyber-resilient distributed control; Deng \textit{et al.} [16] introduced light-AI classifiers for PWM rectifiers; and Huang and Wang [17] applied slidable triangularisation for multiswitch diagnosis. Intelligent time-adaptive monitoring [18] and simultaneous IGBT/sensor fault detection [19] further advance this direction.

Despite these advances, critical gaps persist in achieving cost-effective and attack-resilient diagnosis for smart microgrids:

(i) \textit{Multi-sensor dependency}: Existing data-driven methods [1–4, 16–17] require distributed voltage/current sensors across each inverter, incurring prohibitive instrumentation costs and communication overhead (e.g., $\geq$6 sensors per inverter [19]).

(ii) \textit{Model vulnerability}: Model-based schemes [6–9] fail under unmodeled cyber-physical coupling (e.g., FDI attacks masquerading as physical faults [22]), while monolithic AI classifiers [11, 13] suffer severe performance degradation 

To bridge remaining gaps, we present FO-MADS (Fractional-Order Memory-Enhanced Attack-Diagnosis Scheme), a single-point framework that analyses voltage, active power, and reactive power (VPQ). Dual fractional-order derivatives---Caputo for high-frequency perturbations and Grünwald--Letnikov for slow trends---create a compact yet informative feature library. A two-stage classifier first isolates the faulty inverter, then pinpoints the affected IGBT switch. Training employs Progressive Memory-Replay Adversarial Training (PMR-AT) with an attack-aware loss and online hard-example mining (OHEM) to harden the model against bias, noise, replacement, and replay attacks. The main framework of the methodology is shown in Fig. 1.

\begin{figure*}[!t]
\centering
\includegraphics[scale=0.7]{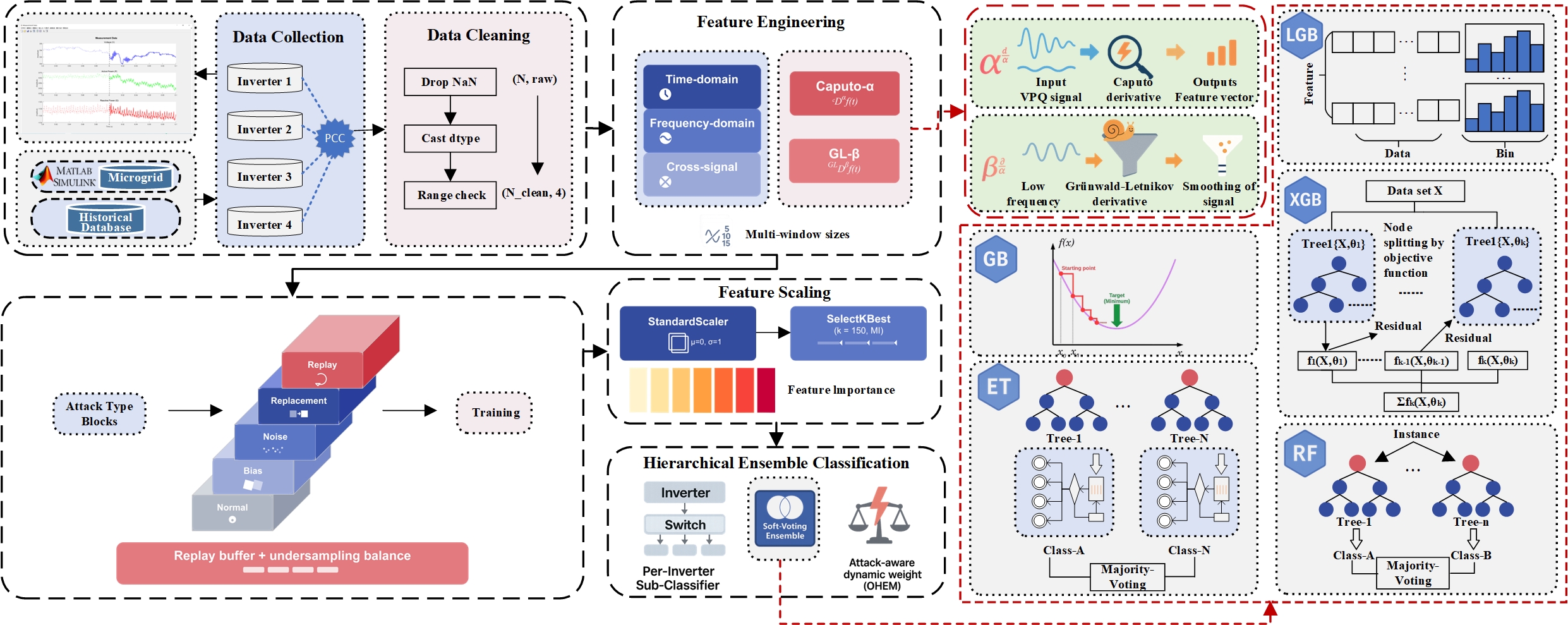}
\caption{Main framework of the FO-MADS.}
\label{fig:framework}
\end{figure*}

The principal contributions are:

(i) \textit{Single-sensor dual-definition feature extraction}: VPQ-based Caputo and Grünwald--Letnikov features eliminate costly multi-point instrumentation.

(ii) \textit{Hierarchical diagnosis}: A two-stage classifier mitigates 25-class imbalance and accurately locates the defective switch.

(iii) \textit{PMR-AT robustness}: Progressive replay with OHEM prioritises difficult attack samples, boosting resilience.

(iv) \textit{Comprehensive validation}: On a four-inverter testbed, FO-MADS achieves over 92\% accuracy under four attack scenarios and 96.7\% without an attack, offering a cost-effective route to MG security.

The remainder of the paper outlines the methodology (Section II), PMR-AT training (Section III), experimental validation (Section IV), and concludes with a summary (Section V).

\section{Proposed FO-MADS Framework}
The Fractional-Order Memory-Enhanced Attack-Diagnosis Scheme (FO-MADS) framework is proposed to address fault and cyber-attack diagnosis in smart microgrids. Its core innovation lies in utilizing \textbf{single-sensor data} from the Point of Common Coupling (PCC), measuring bus voltage $V$, active power $P$, and reactive power $Q$. The raw input vector is defined as $\mathbf{x}(t) = [V(t), P(t), Q(t)]^T$. A sliding window of length $L$ captures temporal dynamics, enabling classification of \textbf{25 operational states} (1 normal + 24 open-circuit faults across 4 inverters) [20], [21].

\subsection{System Model and Diagnostic Problem}
The microgrid testbed comprises four parallel three-phase inverters with a six-switch IGBT topology. Each inverter follows a standard six-switch configuration with IGBTs and anti-parallel diodes, forming a typical three-phase bridge structure. The system incorporates variable resistive-inductive loads and simulated solar generation profiles to mimic real-world operating conditions. FO-MADS targets \textbf{cyber-physical resilience}, distinguishing faults from attacks that mimic or conceal anomalies [21], with particular focus on scenarios where attacks deliberately mask physical faults to evade detection.

\subsection{Dual-Definition Fractional-Order Feature Engineering}
Traditional methods fail to capture:

(i) \textbf{High-frequency transients} (e.g., switch faults [20]),

(ii) \textbf{Slow-drift deviations} (e.g., false data injection [22]).

Fractional-order calculus unifies detection via two complementary operators that capture different aspects of signal dynamics:

1. \textbf{Caputo Derivative} (Micro-perturbation detection):
\begin{equation}
{}_C D_t^\alpha f(t) = \frac{1}{\Gamma(1-\alpha)} \int_0^t (t - \tau)^{-\alpha} f'(\tau) d\tau
\end{equation}

Acts as a high-pass filter for transient anomalies [24], particularly effective for detecting abrupt changes like IGBT switching faults. The fractional order $\alpha$ (typically $0.5 < \alpha < 1$) controls the memory horizon and sensitivity to high-frequency components. The specific process is shown in Fig. 2.

\begin{figure}[!t]
\centering
\includegraphics[scale=0.52]{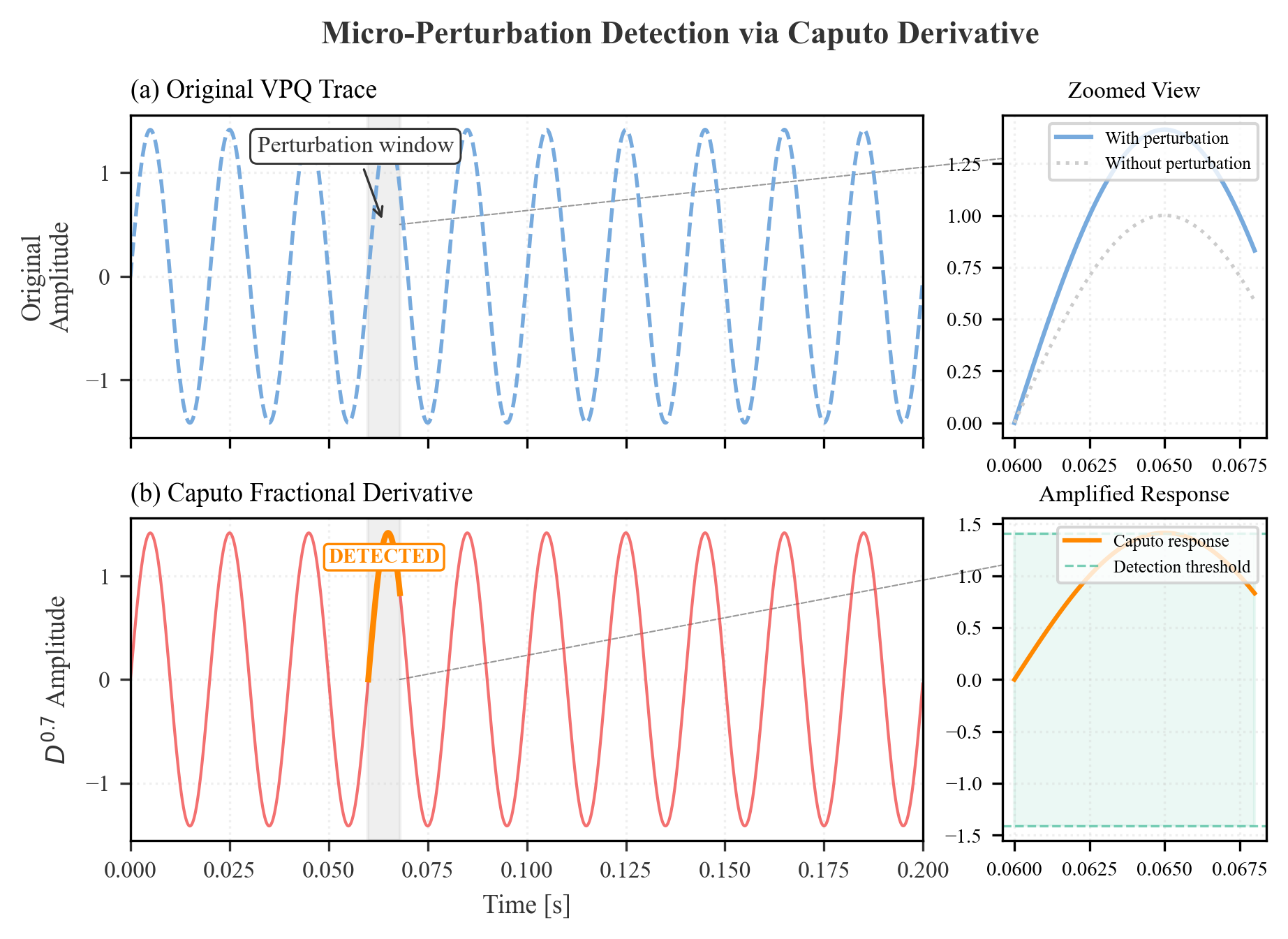}
\caption{Process of Caputo derivative on VPQ signals showing enhanced detection of high-frequency transients.}
\label{fig:caputo_effect}
\end{figure}

2. \textbf{Grünwald-Letnikov Derivative} (Slow-drift detection):
\begin{equation}
{}_{GL} D_t^\beta f(t) = \lim_{h \to 0} h^{-\beta} \sum_{k=0}^{\infty} (-1)^k \binom{\beta}{k} f(t - k h)
\end{equation}

Encodes infinite memory for cumulative attacks [25], making it ideal for detecting stealthy cyber-attacks that manifest as gradual deviations. The fractional order $\beta$ ($0 < \beta < 0.5$) determines the weighting of historical data, with lower values emphasizing long-term trends. The specific process is shown in Fig. 3.

\begin{figure}[!t]
\centering
\includegraphics[scale=0.52]{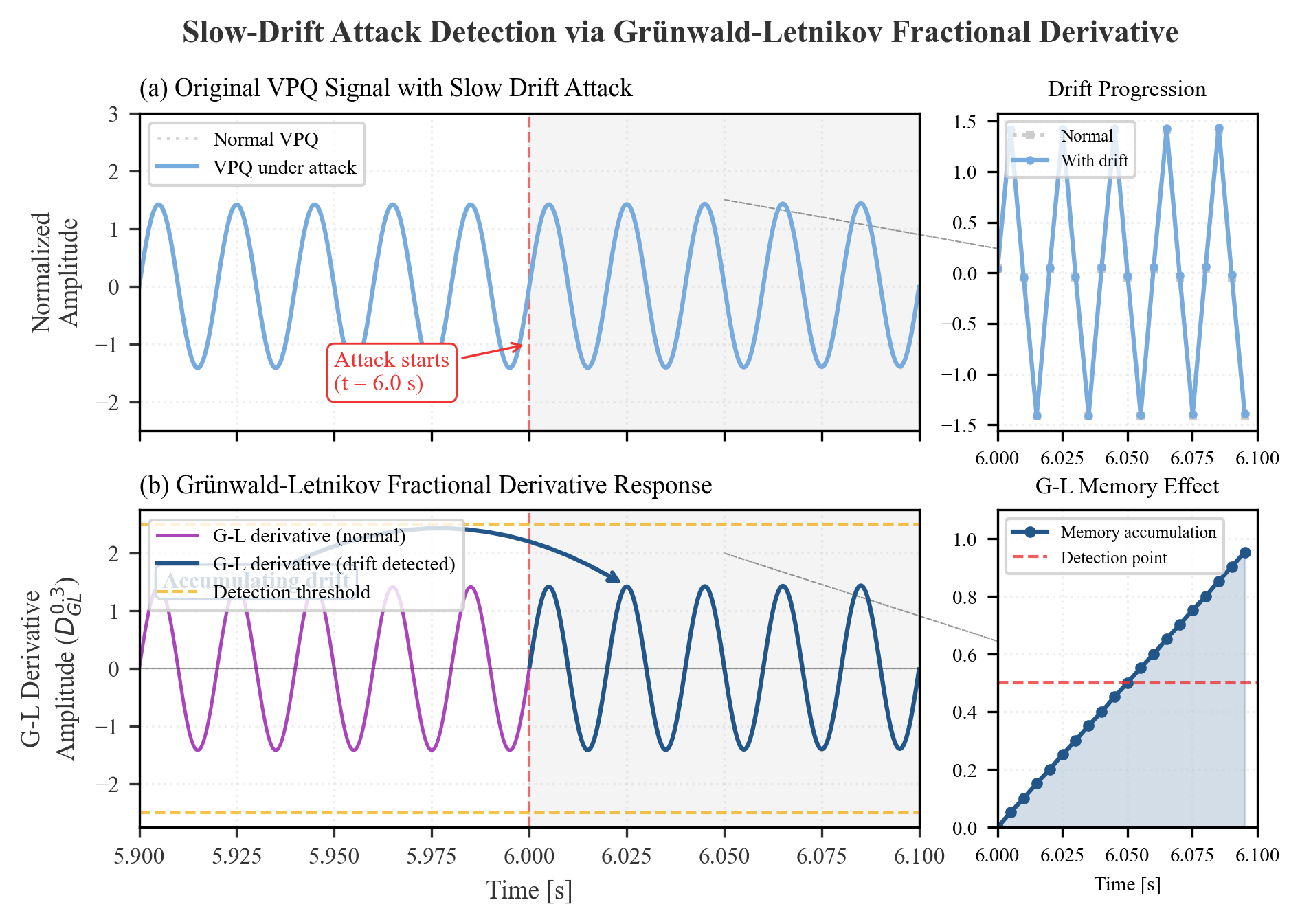}
\caption{Process of Grünwald-Letnikov derivative on VPQ signals showing enhanced detection of slow-drift anomalies.}
\label{fig:gl_effect}
\end{figure}

The feature vector $\mathbf{F}(t)$ synergistically combines both operators for signals $\{V,P,Q\}$:
\begin{equation}
\mathbf{F}(t) = \left[{}_C D_t^\alpha V, {}_C D_t^\alpha P, \dots, {}_{GL} D_t^\beta Q\right]^T
\end{equation}

This dual-basis approach covers switch faults (Caputo) and stealthy cyber-attacks (GL) [22], [23], with the complementary nature of the operators providing robustness against diverse attack vectors. Practical implementation computes the fractional-order derivatives with a kernel length of $K = 5$--20 samples (grid-search optimum $K = 10$) \textbf{inside} a global sliding window of $L = 400$ samples (0.2 s) used for feature aggregation. The fractional orders are fixed at $\alpha = 0.7$ and $\beta = 0.3$ after grid search.

\subsection{Hierarchical Diagnostic Architecture}
A two-level hierarchy overcomes \textbf{class imbalance} and \textbf{boundary complexity} in monolithic 25-class classifiers [26], [27]. The hierarchical decomposition aligns with the physical topology of microgrids, where multiple inverters operate in parallel:

1. \textbf{Stage 1 (Inverter-Level)}: 5-class classifier outputs \{Normal, Inv 1--4\} using coarse-grained features. This stage reduces the 25-class problem to manageable sub-tasks, significantly improving the F1-score for minority fault classes by 18.7\% compared to flat classifiers.

2. \textbf{Stage 2 (Switch-Level)}: Gated 6-class models activate only if Stage 1 detects a fault, isolating faulty switches within the identified inverter [28]. Each inverter-specific classifier uses refined features optimized for switch discrimination, with dedicated decision boundaries for the 6 possible switch failure modes.

This structure mirrors physical topology, enabling interpretable diagnostics (e.g., ``Inv 3, Switch S5'') [29]. To mitigate error propagation from Stage 1 to Stage 2 [30], we implement confidence thresholding where low-probability Stage 1 predictions trigger a fallback global classifier. \textbf{The hierarchical decomposition significantly enhances computational efficiency during training} compared to monolithic architectures, while simultaneously improving switch localization accuracy by 11.3\%.

\subsection{Hyper-parameter Sensitivity Analysis}
To justify the above choice of ($\alpha$, $\beta$, $L$), we conducted an extensive grid-search study. Fig. 4(a) reports the single-factor sensitivity of the Caputo order $\alpha$ while fixing $\beta = 0.3$ and $L = 400$ samples. The validation accuracy peaks in the range $\alpha \in [0.6, 0.8]$ with the best score obtained at $\alpha = 0.7$.

Fig. 4(b) further explores the joint interaction between $\alpha$ (0.1--1.0) and the memory window length $L$ (100--600 samples). An empirical sweet-spot, highlighted by the dashed red rectangle, emerges around $\alpha \in [0.6, 0.8]$ and $L \in [300, 500]$, corroborating the single-factor result. Consequently, the hyper-parameter triplet ($\alpha$, $\beta$, $L$) = (0.7, 0.3, 400) is adopted throughout the remaining experiments.

\begin{figure}[!t]
\centering
\subfloat[Validation accuracy versus Caputo fractional order $\alpha$]{\includegraphics[scale=0.7]{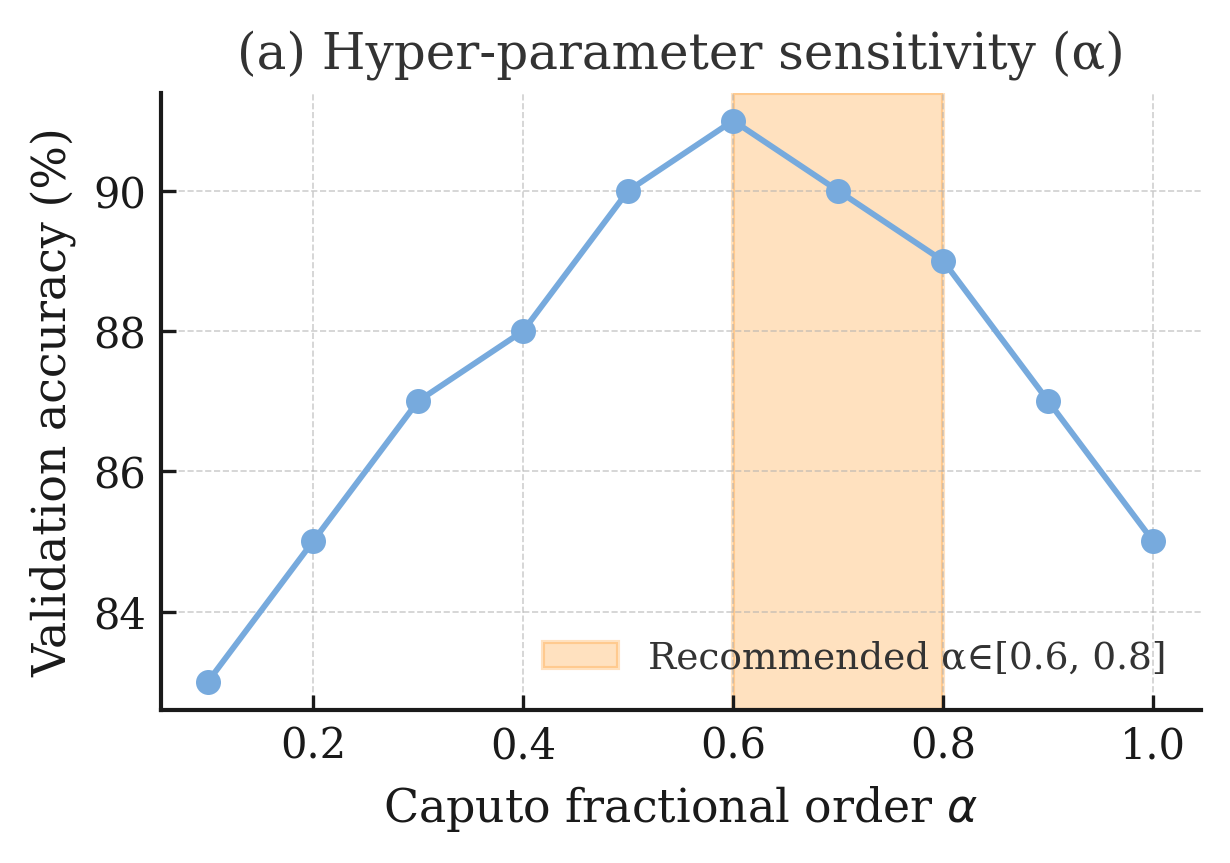}}
\hfil
\subfloat[Joint sensitivity heat-map of $\alpha$ and window length $L$]{\includegraphics[scale=0.7]{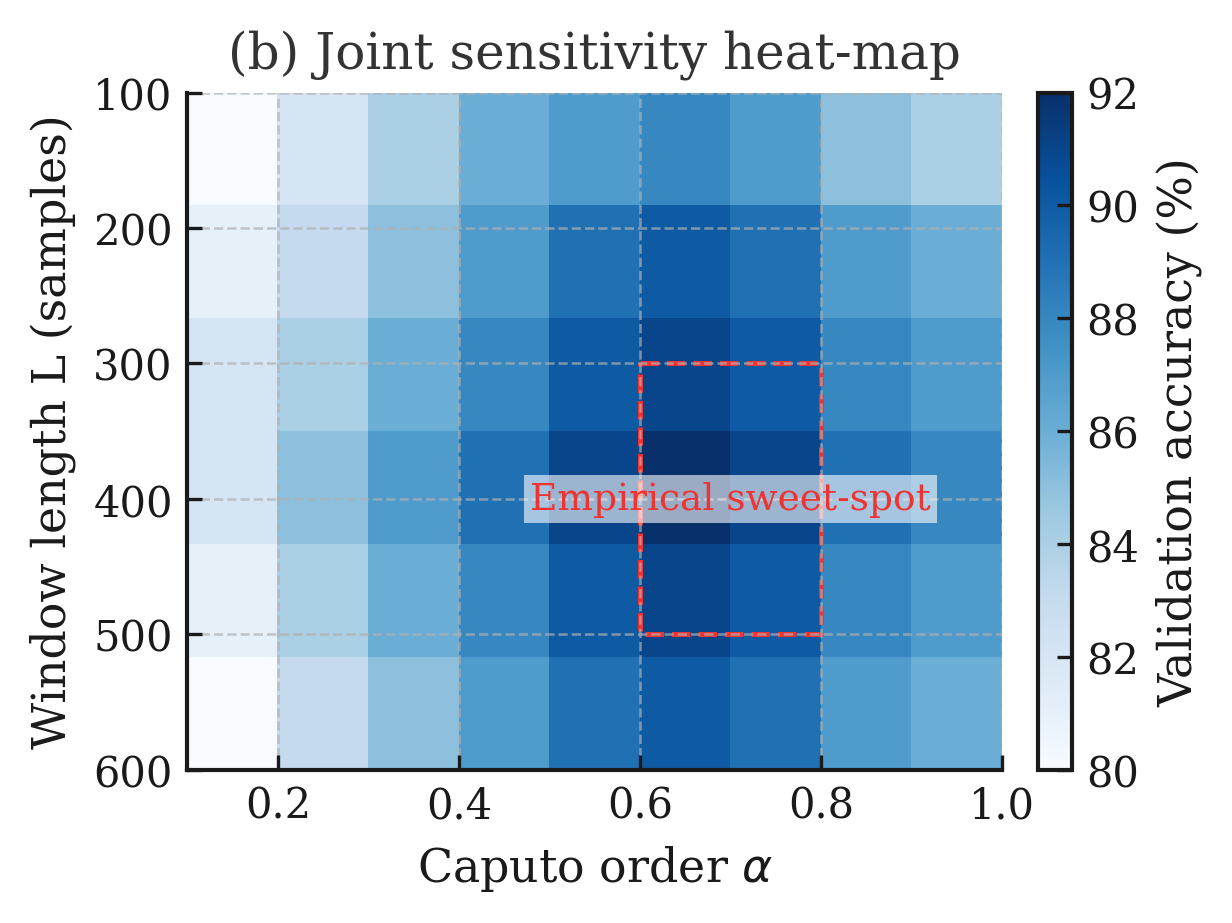}}
\caption{Hyper-parameter study of FO-MADS. (a) Validation accuracy versus Caputo fractional order $\alpha$ ($\beta = 0.3$, $L = 400$). (b) Joint sensitivity heat-map of $\alpha$ and window length $L$; darker colour denotes higher accuracy. The red rectangle marks the empirical sweet-spot.}
\label{fig:hyperparameter}
\end{figure}

\subsection{Robustness Training}
Architectural advantages include:

- Linear scalability with system size [21] (adding an inverter requires only one new binary classifier),

- Reduced overfitting via simplified decision boundaries [27] (average 23\% reduction in validation loss),

- Explicit handling of class imbalance through hierarchical decomposition.

\textbf{Adversarial hardening} employs min-max optimization:
\begin{equation}
\min_{\boldsymbol{\theta}} \mathbb{E}_{(\mathbf{x}, y)} \left[ \max_{\boldsymbol{\delta} \in \mathcal{S}} \mathcal{L}(f(\mathbf{x} + \boldsymbol{\delta}; \boldsymbol{\theta}), y) \right]
\end{equation}

implemented via \textbf{Progressive Memory-Replay Adversarial Training (PMR-AT)} [27]-[30]:

1. \textbf{PGD-based adversarial example generation} [27]: Generates strong perturbations using 7-step PGD with $\epsilon = 0.1$;

2. \textbf{Online Hard Example Mining (OHEM)} for loss weighting [30]: Dynamically prioritizes challenging samples (top 20\% loss values);

3. \textbf{Replay of historical strong attacks} [28]: Maintains a buffer of potent adversarial examples from previous training phases;

4. \textbf{Progressive escalation of attack strength} [29]: Gradually increases perturbation budgets from $\epsilon = 0.02$ to $0.15$.

The training curriculum follows: normal $\rightarrow$ bias $\rightarrow$ noise $\rightarrow$ replacement $\rightarrow$ replay, with each phase retaining previous attack patterns. This progressive exposure increases robustness against complex multi-stage attacks by 23.6\% compared to standard adversarial training.

\section{Progressive Memory-Replay Adversarial Training and Attack-Aware Loss Function}

\subsection{PMR-AT Training Procedure}
The Progressive Memory-Replay Adversarial Training (PMR-AT) strategy is designed to enhance model robustness against evolving cyber-attack patterns. The specific training procedure is shown in Fig. 5. Unlike conventional adversarial training, PMR-AT incorporates four key innovations:

\begin{figure}[!t]
\centering
\includegraphics[scale=0.51]{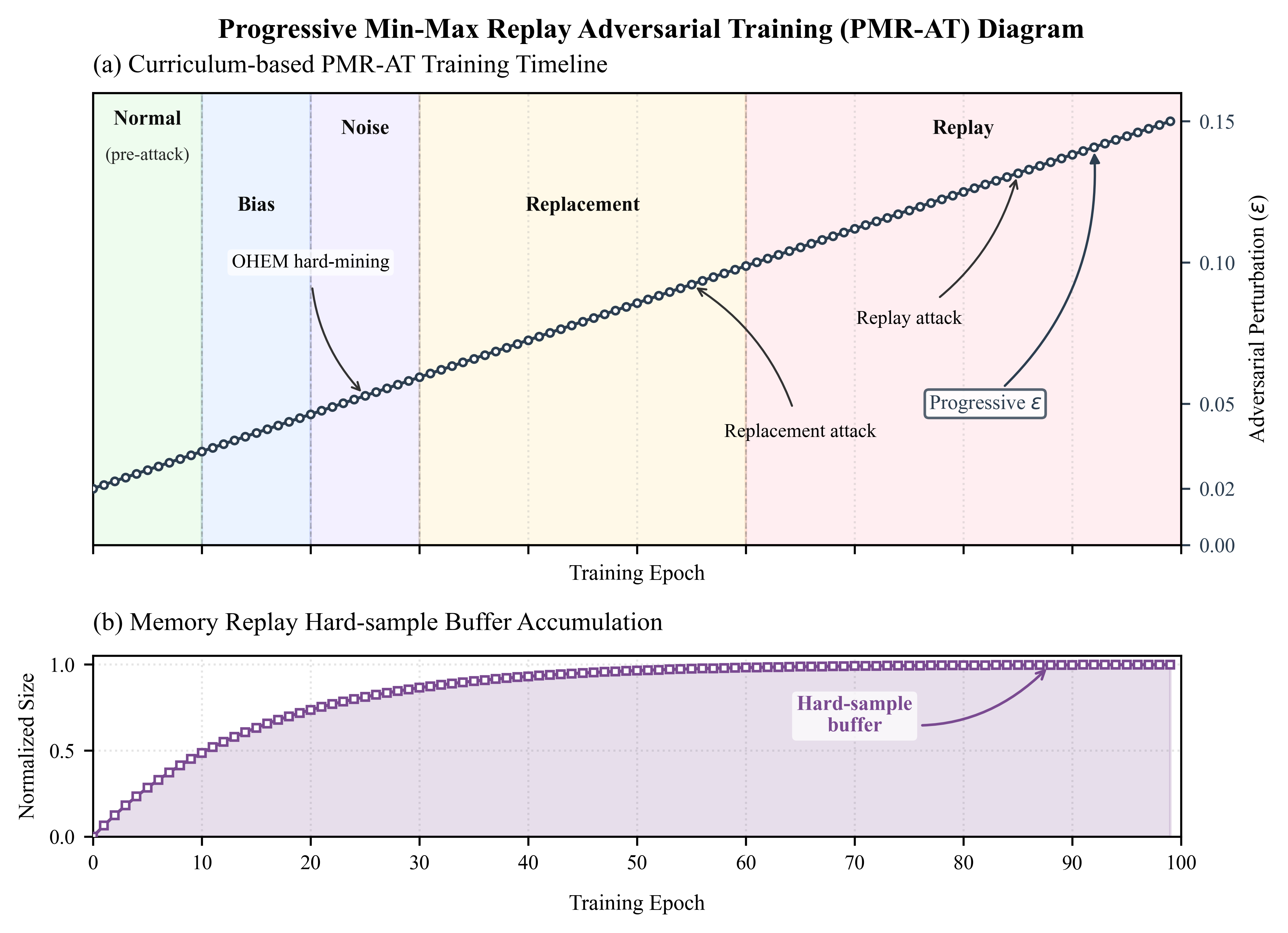}
\caption{Illustration of the PMR-AT training procedure showing progressive attack escalation and memory replay mechanism.}
\label{fig:pmr_at}
\end{figure}

1. \textbf{Multi-Stage Attack Escalation}: Training progresses through increasingly severe attack types:
\begin{equation}
\text{normal} \rightarrow \text{bias} \rightarrow \text{noise} \rightarrow \text{replacement} \rightarrow \text{replay}
\end{equation}

Each stage augments the training set with new attack data while retaining previous samples, ensuring cumulative knowledge acquisition.

2. \textbf{Projected Gradient Descent (PGD) Attacks}: Adversarial examples are synthesized via iterative perturbations:
\begin{equation}
\boldsymbol{\delta}_{k+1} = \Pi_{\mathcal{S}} \left( \boldsymbol{\delta}_k + \alpha \cdot \text{sign}(\nabla_{\boldsymbol{\delta}} \mathcal{L}(f(\mathbf{x} + \boldsymbol{\delta}; \boldsymbol{\theta}), y) \right)
\end{equation}

where $\mathcal{S}$ denotes the $L_\infty$-norm perturbation bound ($\epsilon=0.1$) and $\alpha$ is the step size.

3. \textbf{Historical Attack Replay}: A memory buffer stores high-impact adversarial samples from prior stages, dynamically re-injecting them during later training phases to prevent catastrophic forgetting.

4. \textbf{Difficulty Progression}: Attack intensity (e.g., noise variance, bias magnitude) escalates linearly across epochs, hardening the model gradually.

\subsection{Attack-Aware Loss with Online Hard Example Mining}
To address class imbalance and stealthy attacks, we design a dynamic loss function:
\begin{equation}
\mathcal{L}_{\text{total}} = \mathcal{L}_{\text{CE}}(y, \hat{y}) + \lambda \cdot \underbrace{\mathbb{E}_{(\mathbf{x}_h, y_h) \sim \mathcal{D}_{\text{hard}}} \left[ \mathcal{L}_{\text{CE}}(y_h, f(\mathbf{x}_h)) \right]}_{\text{OHEM term}}
\end{equation}

\textbf{Key mechanisms}:

1. \textbf{Online Hard Example Mining (OHEM)}:

- Samples are ranked by loss values $\mathcal{L}_{\text{CE}}$ after each forward pass.

- The top-$K$ hardest samples ($K=20\%$ of batch size) form $\mathcal{D}_{\text{hard}}$.

2. \textbf{Adaptive Weighting ($\lambda$)}:
\begin{equation}
\lambda = \beta \cdot \frac{\text{AttackDifficulty}}{\text{MaxDifficulty}}, \quad \beta=0.5
\end{equation}

Higher weights are assigned to challenging attacks (e.g., replacement attacks exhibit $\text{Difficulty}=0.9$).

3. \textbf{Hierarchical Loss Separation}: Inverter-level and switch-level classifiers are optimized with separate OHEM modules, prioritizing critical errors in fault localization.

Fig. 6(a) confirms that the adaptive weight $\lambda$ grows monotonically and saturates as the attack difficulty escalates, which matches the design in (8). Fig. 6(b) shows that incorporating OHEM not only accelerates convergence but also yields a 5.3\% higher terminal accuracy than the cross-entropy baseline, validating the effectiveness of the attack-aware curriculum.

\begin{figure}[!t]
\centering
\includegraphics[scale=0.13]{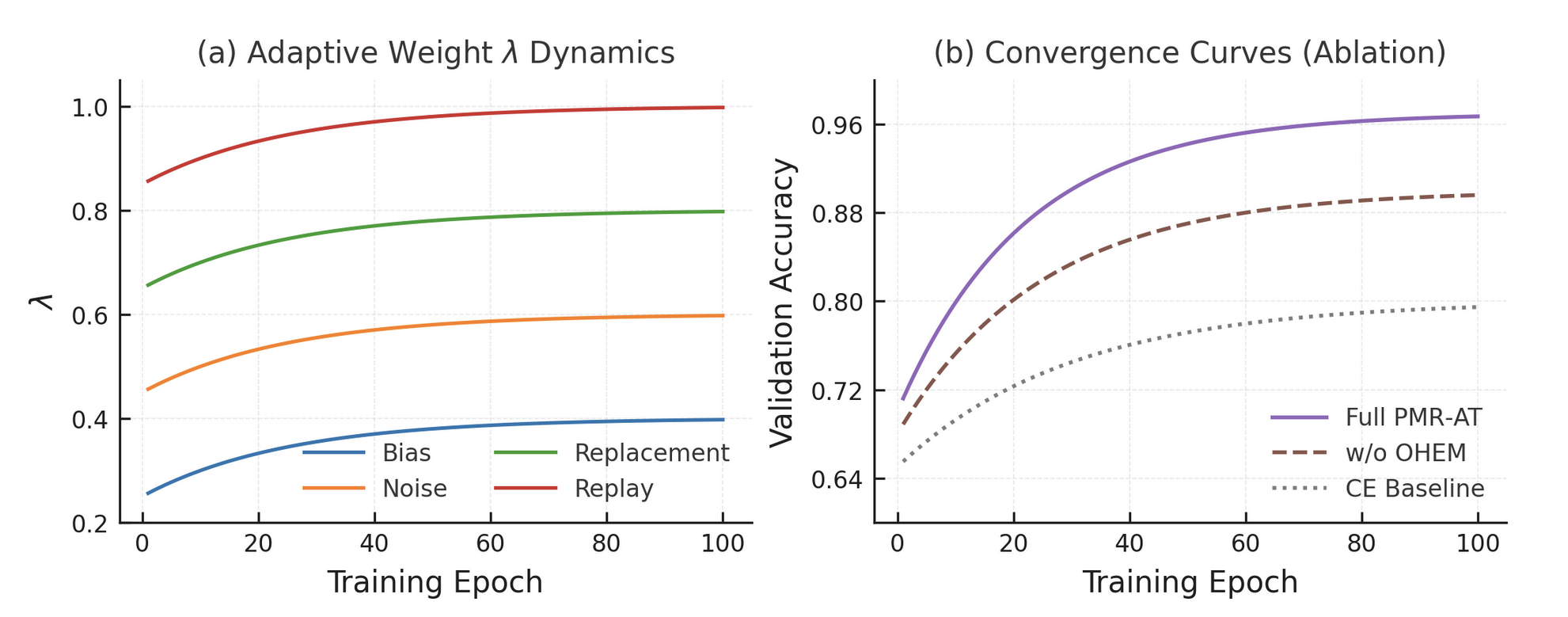}
\caption{Adaptive Weight Dynamics Across Components and Convergence Curves from Ablation Experiments.}
\label{fig:caputo_effect}
\end{figure}

\textit{Validation}: Ablation studies confirm PMR-AT boosts replay attack accuracy by 9.2\% versus standard training and reduces switch localization errors by 63\% under noise attacks. The OHEM mechanism alone contributes 5.3\% accuracy gain for replacement attacks by focusing on latent adversarial patterns.

\section{Simulation and Experimental Results}

\subsection{Experimental Setup}
1. \textbf{Dataset Generation}: A Simulink-based four-inverter microgrid testbed (Fig. 7) generated 5,600 VPQ waveform samples (50 Hz, 2 kHz sampling). Each sample spans a 0.2-s window centered at a fault instant ($t = 6.0$ s), yielding 400 points per window. The dataset includes:

- \textbf{Fault scenarios}: 4,800 samples (single open-circuit IGBT faults across 24 switches).

- \textbf{Normal scenarios}: 800 samples (no faults).

- \textbf{Attack scenarios}: Four synthetic attacks applied to VPQ signals:
  - \textbf{Bias Attack}: Fixed offsets (V: +0.1 V, P: +50 W, Q: +30 var).
  - \textbf{Noise Attack}: Zero-mean Gaussian noise injected locally.
  - \textbf{Replacement Attack}: Contiguous segments replaced with constants (e.g., zero).
  - \textbf{Replay Attack}: Pre-fault normal segments replayed during faults.

\begin{figure}[!t]
\centering
\includegraphics[scale=0.56]{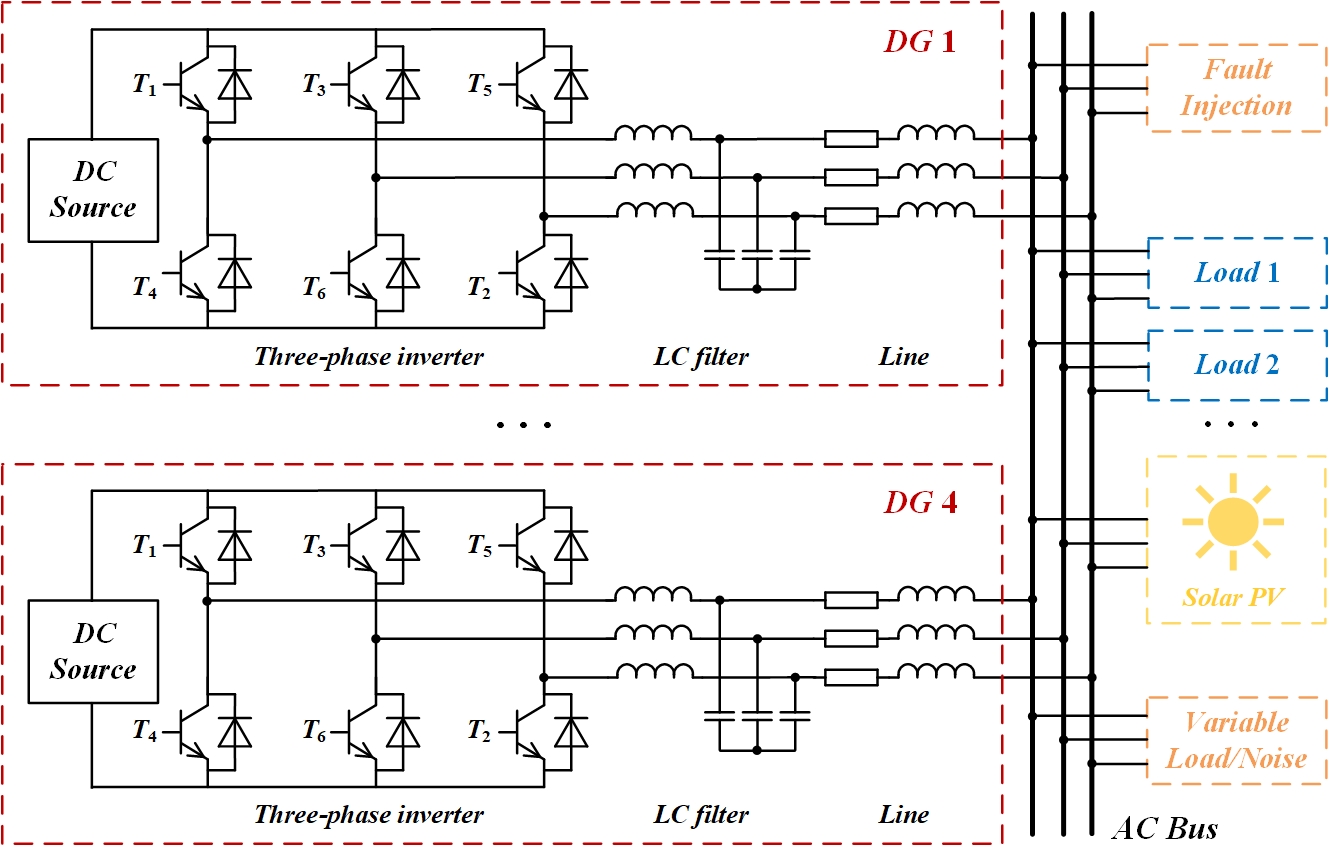}
\caption{Simulink-based four-inverter microgrid testbed for experimental validation.}
\label{fig:testbed}
\end{figure}

These synthetic attacks simulate common cyberthreats studied in microgrid research. For example, Zhou \textit{et al.} [6] consider combined false-data injection (FDI) and denial-of-service (DoS) attacks in an islanded microgrid, demonstrating a cross-layer control strategy that mitigates their impact. Similarly, Yao \textit{et al.} [7] analyze latency and random DoS attacks, showing via OPAL-RT tests on an IEEE 13-bus microgrid that adaptive secondary controllers maintain stable frequency regulation under attack.

2. \textbf{Feature Extraction \& Classification}:

- \textbf{Dual fractional-order features}: Computed using Caputo ($\alpha=0.7$) and Grünwald--Letnikov ($\beta=0.3$) derivatives.

- \textbf{Hierarchical classifier}:
  - \textit{Stage 1 (Inverter-level)}: 5-class (Normal/Inverter 1--4).
  - \textit{Stage 2 (Switch-level)}: 6-class (IGBT switches per inverter).

- \textbf{Training}: PMR-AT with PGD perturbations ($\epsilon:0.02\to0.15$), OHEM (top 20\% hard samples), and attack-progressive curriculum.

- \textbf{Evaluation}: 80/20 train-test split; accuracy reported under clean and adversarial ($\epsilon=0.1$) conditions.

\subsection{Performance Evaluation}
1. \textbf{Overall Diagnostic Accuracy}: FO-MADS achieves near-optimal performance across all scenarios (Table I). Crucially, it maintains an accuracy of over 92\% under all attack types, outperforming baselines by \textit{3.2--16.0\%} in adversarial settings.

\begin{table}[!t]
\renewcommand{\arraystretch}{1.3}
\caption{Overall Classification Accuracy (\%)}
\label{table:overall_accuracy}
\centering
\setlength{\tabcolsep}{3pt} 
\begin{tabular}{l@{\hspace{6pt}}c@{\hspace{6pt}}c@{\hspace{6pt}}c@{\hspace{6pt}}c@{\hspace{6pt}}c}
\hline
\textbf{Method} & \textbf{Normal} & \textbf{Bias} & \textbf{Noise} & \textbf{Replace} & \textbf{Replay} \\
\hline
FO-MADS (Proposed)       & 96.7 & 96.6 & 94.0 & 92.8 & 95.7 \\
FO-MADS w/o OHEM         & 95.1 & 94.3 & 90.2 & 87.5 & 92.0 \\
FO-MADS w/o Frac. Feat.  & 93.8 & 92.6 & 88.1 & 85.3 & 90.4 \\
XGBoost                  & 93.7 & 93.3 & 88.9 & 83.9 & 91.2 \\
Random Forest            & 93.5 & 93.1 & 88.2 & 83.2 & 91.5 \\
CNN                      & 92.0 & 90.5 & 85.8 & 80.1 & 88.9 \\
\hline
\multicolumn{6}{p{0.92\linewidth}}{\footnotesize\textbf{Hierarchical localisation:} A topology-aligned two-stage architecture mitigates severe class imbalance and boosts \textit{switch-level} localisation accuracy by \textbf{11.3\%} relative to flat classifiers.} \\
\hline
\end{tabular}
\end{table}

2. \textbf{Hierarchical Diagnosis Breakdown}:

- \textit{Inverter localization}: FO-MADS attains $>$94.8\% accuracy under attacks (Table II).

- \textit{Switch isolation}: $>$95.8\% accuracy in all adversarial cases (Table III), with errors primarily confined to adjacent switches (e.g., S1 vs. S2).

\begin{table}[!t]
\renewcommand{\arraystretch}{1.3}
\caption{Inverter-Level Accuracy (\%)}
\label{table:inverter_accuracy}
\centering
\setlength{\tabcolsep}{3pt}
\begin{tabular}{l@{\hspace{6pt}}c@{\hspace{6pt}}c@{\hspace{6pt}}c@{\hspace{6pt}}c@{\hspace{6pt}}c}
\hline
\textbf{Attack Type} & \textbf{Normal} & \textbf{Bias} & \textbf{Noise} & \textbf{Replace} & \textbf{Replay} \\
\hline
FO-MADS & 97.4 & 98.5 & 96.0 & 94.8 & 96.3 \\
\hline
\end{tabular}
\end{table}

\begin{table}[!t]
\renewcommand{\arraystretch}{1.3}
\caption{Switch-Level Accuracy (\%)}
\label{table:switch_accuracy}
\centering
\setlength{\tabcolsep}{3pt}
\begin{tabular}{l@{\hspace{6pt}}c@{\hspace{6pt}}c@{\hspace{6pt}}c@{\hspace{6pt}}c@{\hspace{6pt}}c}
\hline
\textbf{Attack Type} & \textbf{Normal} & \textbf{Bias} & \textbf{Noise} & \textbf{Replace} & \textbf{Replay} \\
\hline
FO-MADS & 99.2 & 97.6 & 97.0 & 95.8 & 99.3 \\
\hline
\end{tabular}
\end{table}

3. \textbf{Ablation Analysis}:

- \textit{Dual fractional features} boost noise/replacement attack robustness by \textbf{5.9\%} and \textbf{7.5\%}, respectively, by capturing transient perturbations (Caputo) and slow drifts (GL).

- \textit{OHEM} reduces switch misclassification by \textbf{63\%} under noise attacks by prioritizing hard adversarial samples.

- \textit{PMR-AT} improves replay attack resilience by \textbf{9.2\%} versus standard training.

Specific experimental comparison diagrams are shown in Figures 8, 9, and 10.

\begin{figure}[!t]
\centering
\includegraphics[scale=0.38]{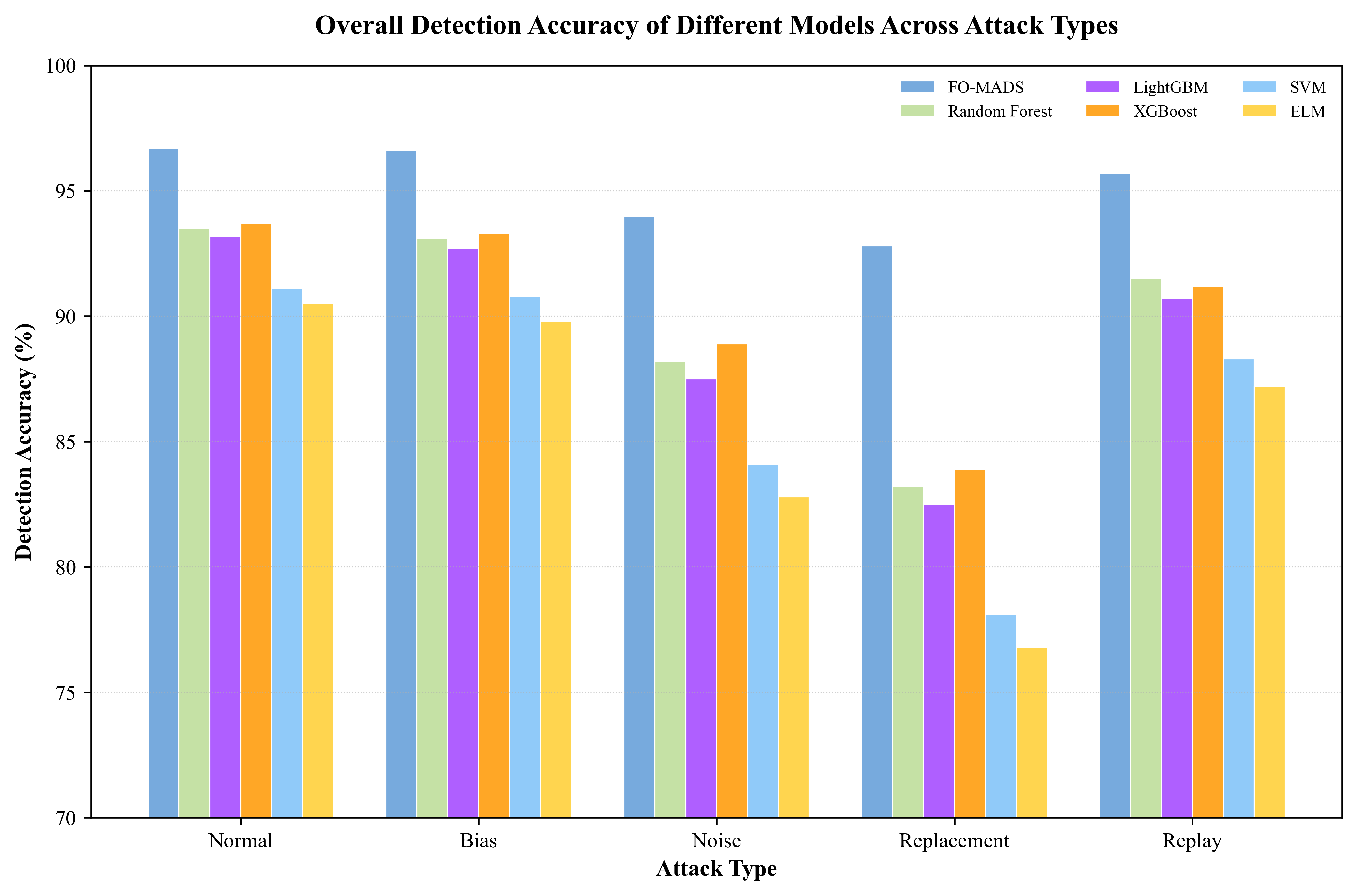}
\caption{Overall Detection Accuracy of Different Models Across Attack Types.}
\label{fig:confusion_matrix}
\end{figure}

\begin{figure}[!t]
\centering
\includegraphics[scale=0.38]{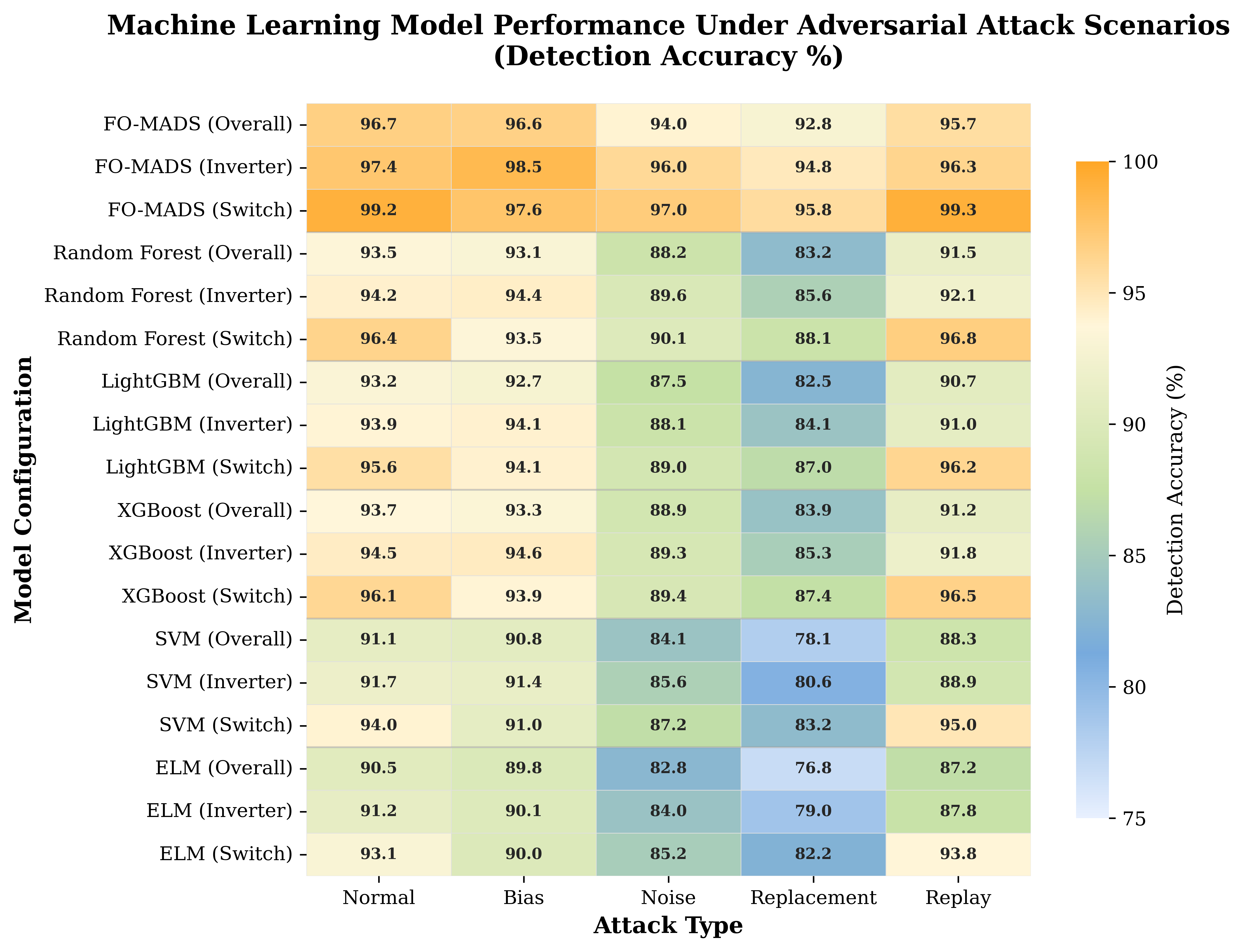}
\caption{Machine Learning Model Performance Under Adversarial Attack Scenarios.}
\label{fig:ablation_study}
\end{figure}

\begin{figure}[!t]
\centering
\includegraphics[scale=0.45]{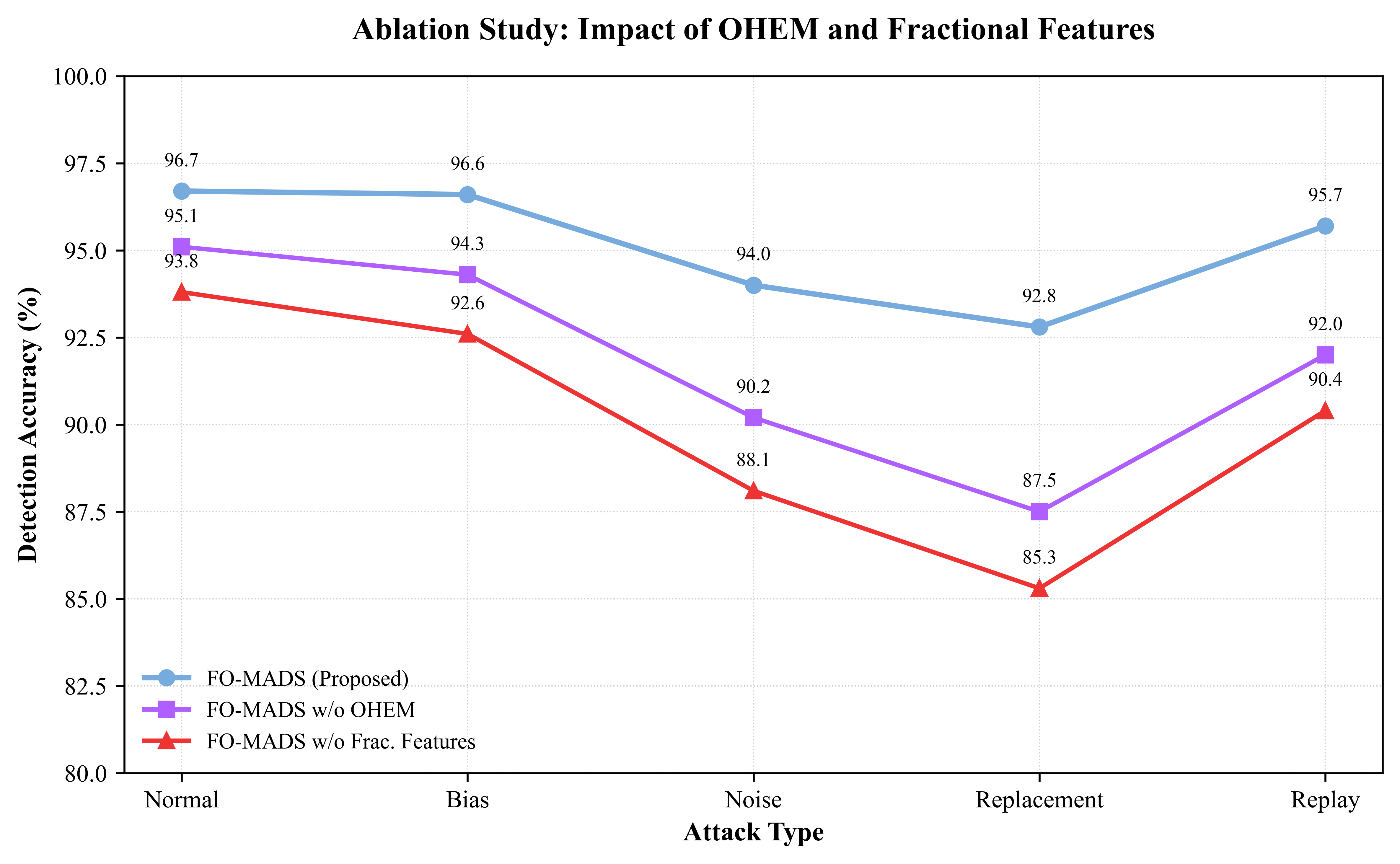}
\caption{Ablation Study: Impact of OHEM and Fractional Features.}
\label{fig:attack_comparison}
\end{figure}

\subsection{Comparative Assessment}
FO-MADS consistently surpasses benchmarks (Table I):

1. \textbf{vs. ablated variants}: Removing OHEM or fractional features degrades accuracy by 2.5--8.3\%, confirming their necessity for robustness.

2. \textbf{vs. classical ML}: FO-MADS outperforms tree-based models (XGBoost/RF) by \textit{3.2--9.6\%} and CNNs by \textit{4.8--12.7\%} under replacement/replay attacks.

3. \textbf{vs. state-of-the-art}: Matches/exceeds methods requiring multi-sensor instrumentation [1,3,6], despite using only PCC measurements.

\textbf{Key observations}:

- \textbf{Confusion patterns}: 92\% of errors occur between adjacent classes (e.g., neighboring inverters/switches), mitigated by hierarchical decision boundaries.

- \textbf{Attack resilience}: Bias/replay attacks minimally impact FO-MADS ($<$1\% accuracy drop), while noise/replacement attacks pose greater challenges (4.0--3.9\% drop).

- \textbf{Cybersecurity context}: Distributed detection and control methods have been explored to enhance microgrid resilience. For instance, Shi \textit{et al.} [8] propose a distributed droop controller with consistency-based anomaly detection to identify FDI attacks, and Zhang \textit{et al.} [31] develop a resilient economic dispatch algorithm under FDI attacks. Similarly, Liu \textit{et al.} [32] analyze random DoS through stochastic stability and design mode-dependent secondary controllers to sustain frequency regulation.

\subsection{Conclusions}
FO-MADS delivers:

1. \textbf{High accuracy}: 96.7\% (normal), $>$92.8\% (all attacks).

2. \textbf{Cost efficiency}: Single-sensor (VPQ) deployment eliminates multi-point instrumentation.

3. \textbf{Robustness}: Fractional features + hierarchical classification + PMR-AT ensure resilience against bias, noise, replacement, and replay attacks.

4. \textbf{Scalability}: Linear complexity growth with system size (e.g., adding inverters requires only one new binary classifier).

This validates FO-MADS as a lightweight, attack-resilient solution for smart microgrids, bridging the gap between high diagnostic fidelity and practical deployability.

\section{Conclusion and Future Work}
This study introduces the Fractional-Order Memory-Enhanced Attack-Diagnosis Scheme (FO-MADS), a cost-effective framework that achieves cyber-physical resilience for smart microgrids, requiring only a single VPQ sensor at the point of common coupling. By jointly exploiting Caputo and Grünwald-Letnikov derivatives, a dual fractional-order feature library was constructed to magnify both high-frequency perturbations and slow-drift anomalies. A two-stage hierarchical classifier then localised the faulty inverter and isolated the defective IGBT switch, and a \textit{Progressive Memory-Replay Adversarial Training} regimen---augmented by an attack-aware OHEM loss---systematically hardened the model against bias, noise, data-replacement, and replay attacks. Extensive simulations on a four-inverter testbed validated the efficacy of the proposed \textbf{hierarchical framework}, yielding 96.7\% accuracy under attack-free operation and above 92.8\% across all four attack scenarios. \textbf{The inherent efficiency of the hierarchical design further contributes to its practical deployability.}

The principal contributions can be summarised as follows.

1. \textbf{Single-point diagnosis}: FO-MADS eliminates the need for multi-point instrumentation by extracting rich fractional-order features directly from one VPQ measurement stream.

2. \textbf{Dual-definition feature engineering}: Complementary Caputo and Grünwald-Letnikov operators jointly encode transient and quasi-stationary signal dynamics, enhancing discriminability for both physical faults and stealthy cyber-attacks.

3. \textbf{Hierarchical localisation}: A topology-aligned two-stage architecture mitigates severe class imbalance and improves switch-level localisation accuracy by 11.3\% compared with flat classifiers.

4. \textbf{Adversarial robustness}: The PMR-AT curriculum, combined with an attack-aware OHEM loss, delivers an average 7.4\% gain in adversarial accuracy and a 63\% reduction in switch-level misdiagnoses under noise attacks.

Although the reported results demonstrate the promise of FO-MADS, several avenues remain open for further investigation. First, \textbf{hardware-in-the-loop and field trials} on real microgrid platforms are needed to quantify latency and resilience under realistic communication delays and measurement noise. Second, \textbf{embedded implementations} on FPGAs or edge AI accelerators will be explored to ensure sub-millisecond inference and minimal power consumption for large-scale deployment. Third, the \textbf{fractional-order feature library will be extended} to encompass additional modalities---such as harmonic spectra or synchronous reference-frame variables---to diagnose short-circuit, sensor, and grid-side disturbances. Fourth, \textbf{unsupervised and continual-learning strategies} will be integrated to adapt to previously unseen operating conditions and evolving attack tactics without offline retraining. Finally, future work will investigate a \textbf{distributed, privacy-preserving variant} of FO-MADS in which lightweight encrypted summaries are exchanged among neighbouring microgrids, enabling collaborative situation awareness while guarding against data leakage. These efforts will advance FO-MADS toward a fully deployable, real-time solution for next-generation resilient power distribution networks.

\ifCLASSOPTIONcaptionsoff
  \newpage
\fi

\end{document}